\documentclass[nofootinbib,preprintnumbers,amsmath,amssymb,bibnotes]{revtex4}
 
\usepackage{graphicx}
\usepackage{dcolumn}
\usepackage{bm}
\usepackage{cancel} 

\usepackage{float}
\restylefloat{figure}
\usepackage{epstopdf}

\usepackage{tikz}
\usetikzlibrary{calc} \usetikzlibrary{patterns} \usetikzlibrary{decorations.pathreplacing} \usetikzlibrary{decorations.markings} \usetikzlibrary{decorations.pathmorphing} \usetikzlibrary{positioning}

\newcommand{\be}{\begin{equation}}
\newcommand{\ee}{\end{equation}}
\newcommand{\comment}[1]{}

\newcommand{\req}[1]{(\ref{#1})}

\newcommand{\nwc}{\newcommand}
\nwc{\ba}  {\begin{array}}
\nwc{\ea}  {\end{array}}

\nwc{\bea} {\begin{eqnarray}}
\nwc{\eea} {\end{eqnarray}}
\nwc{\nn} {\nonumber}
\nwc{\nnn} {\nonumber \vspace{.2cm} \\ }

\nwc{\bda} {\bdm\ba{lcl}}
\nwc{\eda} {\ea\edm}

\nwc{\ds}  {\displaystyle}

\nwc{\ra}{\rightarrow}
\nwc{\lra}{\longrightarrow}

\nwc{\p} {\partial}
\nwc{\Tr}{{\rm Tr}}

\def\l{{\lambda}}

\def\g{{\gamma}}

\def\half{{1\over 2}}
\def\p{{\partial}}

\def\t{{\theta}}

\def\({\left(}
\def\){\right)}

\begin{document}

\preprint{AEI--2010--180, MPP--2010--173}

\title{A recursive method for SYM $\bm{n}$--point tree amplitudes}


\author{Carlos R. Mafra$^a$,
Oliver Schlotterer$^b$,
Stephan Stieberger$^b$, and Dimitrios Tsimpis$^c$}
\affiliation{$a$ Max--Planck--Institut f\"ur Gravitationsphysik, Albert--Einstein--Institut,
14476 Potsdam, Germany,}
\affiliation{$^b$ Max--Planck--Institut f\"ur Physik, Werner--Heisenberg--Institut,
80805 M\"unchen, Germany,}
\affiliation{$^c$ Universit\'e Lyon 1, Institut de Physique Nucl\'eaire de Lyon,
69622 Villeurbanne, France}



\begin{abstract}
We present a recursive method for super Yang--Mills color--ordered $n$--point 
tree amplitudes based on the cohomology of pure spinor superspace
in ten space--time dimensions.
The amplitudes are organized into BRST covariant building blocks with diagrammatic interpretation.
Manifestly cyclic expressions (no longer than one line each) are explicitly
given up to $n=10$ and higher leg generalizations are straightforward.

\end{abstract}

\maketitle

\section{Introduction}

Elementary particle physics relies on
the computation of scattering amplitudes in Yang-Mills theory. Parke and Taylor 
found compact and simple expressions for maximally helicity violating (MHV) 
amplitudes in four space-time dimensions \cite{mhv}, which provide an important 
milestone in discovering hidden structures underlying the S--matrix. Many 
formal as well as phenomenological advances followed since then, see
\cite{CachazoGA,BernDW} for some reviews.

Supersymmetric field theories emerge in the low--energy limit of superstring theory, that is why the latter can be used as a 
powerful tool to gain further insights into field theories, see \cite{mss} for a recent example. There are several 
descriptions for the superstring's world--sheet degrees of freedom, and the pure spinor formalism \cite{psf} is the 
only manifestly supersymmetric formulation known so far which can still be quantized covariantly.

In this letter we use the framework of the pure spinor formalism to 
reduce the computation of $n$--point tree amplitudes in ten-dimensional ${\cal N}=1$ super-Yang-Mills theory (SYM) 
to a recursive cohomology problem in pure spinor 
superspace. The end result is the compact formula \req{Npts} for the supersymmetric color-ordered
$n$--point scattering amplitude at tree level.

Up to now, cohomology arguments have been used to propose SYM amplitudes up to seven--point \cite{FTAmps}, and they 
have been identified as the low energy limit of superstring amplitudes up to six--point \cite{MSST}. The main idea 
of \cite{FTAmps} and this article is to bypass taking the field theory limit of a superstring computation and to 
instead fix SYM amplitudes using the BRST cohomology. This is achieved for any number $n$ of external legs in this letter.

Although the pure spinor framework is initially 
adapted to ten space-time dimensions, one can still dimensionally reduce the 
results and extract the physics from any lower dimensional point of view. At 
any rate, the striking simplicity of our results is exhibited without the need
of four--dimensional spinor helicity formalism. Moreover, the simplicity 
is furnished both for MHV and non-MHV helicity configurations in four space--time dimensions.

\section{Pure spinor cohomology formula for $\bm{A_n}$}

The color-ordered tree-level massless super-Yang-Mills amplitudes in ten dimensions will be argued to be 
determined by the pure spinor superspace cohomology formula\footnote{The $n$--point color-ordered 
formul{\ae} in this letter are all for the
  ordering $1,2,{\ldots},n$.},
\be\label{Npts}
{\cal A}_n = \langle E_{i_1{\ldots} i_{n-1}} V_n \rangle\ ,
\ee
where $V_n$ is the vertex operator for the SYM multiplet in the pure spinor approach to superstring theory. 
The bosonic superfields $E_{i_1{\ldots} i_p}$ are closed under the pure spinor BRST charge $Q$ but 
not BRST-exact in the momentum phase space of an $n$--point massless amplitude where the Mandelstam 
variables $s_{i_1{\ldots} i_p} = \half (k_{i_1} + {\ldots} + k_{i_p})^2$ encompassing $n-1$ momenta vanish, $s_{i_1{\ldots} i_{n-1}} = 0$:
\be\label{descent}
QE_{i_1{\ldots} i_p}= 0, \quad E_{i_1{\ldots} i_p} = QM_{i_1{\ldots} i_p} \ {\rm if} \ s_{i_1{\ldots} i_p} \neq 0\ .
\ee
The $\langle {\ldots}  \rangle$ bracket
denotes a zero mode integration prescription automated in \cite{PSS} which extracts the superfield 
components from the enclosed superfields \cite{psf}. More precisely, nonvanishing contributions 
arise from tensor structures of order $\l^3\t^5$ where $\l$ is the ghost variable of the pure 
spinor formalism and $\theta$ the Grassmann odd superspace variable of ten-dimensional ${\cal N}=1$ SYM.

\subsection{BRST building blocks}
\label{SUB23}

The first step in constructing the BRST cohomological objects $E_{i_1{\ldots} i_{n-1}}$ in (\ref{Npts}) is guided by the worldsheet conformal field theory (CFT) of superstring theory in its pure spinor formulation. Apart from the unintegrated vertex operator $V^i = \l^\alpha A_\alpha^i$, the massless level of the BRST cohomology contains the integral over $U^j = \partial \theta^{\alpha} A_\alpha^j + \Pi^m A_m^j + d_\alpha W^{\alpha }_j + \frac{1}{2} N^{mn} {\cal F}^j_{mn}$ along the worldsheet boundary. The so-called integrated vertex operator $U^j$ is built from $h=1$ fields $[\partial \theta^{\alpha} , \Pi^m , d_\alpha,N^{mn} ]$ of the pure spinor CFT contracted with SYM superfields $[A_\alpha^j, A_m^j, W^{\alpha }_j, {\cal F}^j_{mn}]$.

Computing scattering amplitudes involves the residues $L_{2131\ldots p1}$ of the operator product expansion (OPE) of $p-1$ integrated 
vertex operators $U^j(z_j)$ approach their unintegrated counterpart $V^i(z_i)$:
\be \label{OPEs} 
\lim \limits_{z_2\to z_1} V^1(z_1)U^2(z_2)  {L_{21}\over z_{21}}\ , \ \ \ \lim\limits_{z_p \to z_1} L_{2131\ldots(p-1)1}(z_1) U^p(z_p) \rightarrow 
{L_{2131\ldots(p-1)1p1(z_1)}\over z_{p1} } \ .
\ee
Using the explicit form of $V^i$, $U^j$ in terms of SYM superfields and their OPEs we find
$$\ba{lcl}
L_{21} &=& - A^1_m (\l\g^m W^2) - V^1(k^1\cdot A^2)\\[1mm]
L_{2131} &=& - L_{21}((k^1+k^2)\cdot A^3)
+  (\l\g^{m}W^3)  \big[  A^1_m(k^1\cdot A^2) +  A^{1\,n}{\cal F}^2_{mn}
       - (W^1\g_m W^2)\big]
\ea$$
for two and three legs respectively.

The $p$-leg residues $L_{2131\ldots p1}$ by themselves do transform BRST covariantly, e.g.
$$\ba{lcl}
Q L_{ji} &=& s_{ij} V_i V_j\ , \\
Q L_{jiki} &=& s_{ijk} L_{ji}V_k - s_{ij}\big[ L_{kj} V_i - L_{ki}V_j + L_{ji}V_k\big] ,
\ea$$
but they do not exhibit any symmetry properties in the labels $i$, $j$, $k$ 
as required for a diagrammatic interpretation. However, many irreducibles of the symmetric 
group turn out to be BRST exact, e.g. $Q (A_i \cdot A_j) = -2 L_{(ij)}$. 
Only truly BRST cohomological pieces are kept,
$$
T_{ij} := L_{[ji]} = L_{ji} - L_{(ji)} = L_{ji} + {1\over2} Q (A_i \cdot A_j).
$$
Any higher rank residue $L_{21\ldots p1}$ with $p \geq 3$ requires a redefinition in two steps to 
form the so-called BRST building blocks $T_{12\ldots p}$ which ultimately enter the $n$--point SYM amplitude (\ref{Npts}):
$L_{2131\ldots p1} \longrightarrow \tilde T_{123\ldots p} \longrightarrow T_{123\ldots p}$.
A first step $\tilde T_{123\ldots p} = L_{2131\ldots p1} + \ldots$ removes the BRST trivial parts in $Q\tilde T_{123\ldots p}$, e.g.
$$\ba{lcl}
{\tilde T}_{ijk} &\equiv& L_{jiki} + {s_{ij}\over 2}\big[
(A_j \cdot A_k)V_i - (A_i \cdot A_k)V_j + (A_i \cdot A_j)V_k\big]- {s_{ijk}\over 2}(A_i \cdot A_j)V_k\\[1mm]
Q{\tilde T}_{ijk} &=& s_{ijk} T_{ij}V_k - s_{ij}\big[ T_{jk} V_i - T_{ik}V_j + T_{ij}V_k\big]
\ea$$
such that the BRST variation of $\tilde T_{123\ldots p}$ involves $T_{i_1\ldots i_{q<p}}$ rather than $L_{i_2 i_1 \ldots  i_{q<p} i_1}$. 
But there will be BRST exact components in $\tilde T_{123\ldots p}$ which still have to be subtracted
in a second step. For example, there exist superfields $R^{(l)}_{ijk}$ such that 
\cite{MSST,WIP}
$$
 QR_{ijk}^{(1)} = 2\tilde T_{(ij)k}, \quad Q R^{(2)}_{ijk} = 3\tilde T_{[ijk]}.
$$
The following redefinition yields the hook Young tableau $T_{ijk} = T_{[ij]k}$ with $T_{[ijk]} = 0$
$$
T_{ijk} = \tilde T_{ijk} - {1\over2} Q R^{(1)}_{ijk} - {1 \over 3} Q R^{(2)}_{ijk}
$$
suitable to represent field theory diagrams made of cubic vertices. Similarly, one has to remove $p-1$ BRST trivial 
irreducibles from $T_{12 \ldots p} = \tilde T_{12\ldots p} + \ldots$ where the higher order generalizations of
$A_i \cdot A_j$, and $R^{(l)}_{ijk}$ superfields are related to $z_{ij}$ double poles in the OPE of $U^i(z_i) U^j(z_j)$.

The explicit construction of BRST building blocks $T_{12 \ldots p}$ with higher rank $p$ involves two completely straightforward steps: The residue $L_{2131\ldots p1}$ is determined by the OPEs of the conformal worldsheet fields, and the corresponding $\tilde T_{12\ldots p}$ follows from replacing the lower rank $L_{2131\ldots q1} \mapsto T_{12\ldots q}, \ q<p$ within $QL_{2131\ldots p1}$. Only the last step of finding ``parent superfields'' $R^{(i)}_{12\ldots p}$ whose $Q$ variation yields the BRST exact components of $\tilde T_{12\ldots p}$ requires some intuition. We have worked out such higher order generalizations of the $R^{(1)}_{ijk}$ and $R^{(2)}_{ijk}$ above up to $p=5$ (see an appendix of \cite{WIP}) on the basis of a ``trial and error'' analysis.

More generally, each $T_{i_1 \ldots i_p}$ inherits all the symmetries 
of $T_{i_1 \ldots i_{p-1}}$ 
in the first $p-1$ labels, so there is one new identity at each rank $p$ 
(such as $T_{12[34]} + T_{34[12]} = 0$ at $p=4$) which cannot be inferred 
from lower order relatives. It can be determined from the symmetries of the 
diagrams described by $T_{i_1\ldots i_p}$, e.g.
\be\label{Tfiveid}
T_{ijklm} - T_{ijkml}  + T_{lmijk} - T_{lmjik} - T_{lmkij} + T_{lmkji} = 0
\ee
at $p=5$. 
Higher order generalizations of \req{Tfiveid}\ will be listed in \cite{WIP}.

Just like the OPE residues $L_{2131\ldots p1}$ defined by (\ref{OPEs}), the BRST building blocks $T_{12\ldots p}$ transform covariantly under the BRST charge,
\be\label{QTijkl}
\ba{lcl}
Q T_{ijk} &=& s_{ijk} T_{ij}V_k - s_{ij} (T_{ij}V_{k} + T_{jk} V_i + T_{ki} V_j)\\
QT_{ijkl} &=& s_{ijkl}  T_{ijk}  V_l+  s_{ijk}  \(T_{ijl}  V_k-T_{ijk}V_l +T_{ij}T_{kl} \)
\\[1mm] 
&+& s_{ij}  (  V_i  T_{jkl} +  T_{ikl}  V_j  -  T_{ijl}  V_k+  T_{ik}  T_{jl} +  T_{il}  T_{jk}  -  T_{ij}  T_{kl})\ ,
\ea
\ee
once again, we refer the reader to \cite{WIP} for higher order generalizations.

\subsection{Feynman diagrams and Berends-Giele currents}
\label{SUB22}

In this subsection, we give a diagrammatic interpretation of the BRST building blocks $T_{12..p}$ and combine them 
to color ordered field theory amplitudes with one off-shell leg, so-called Berends-Giele currents \cite{Berends:1987me}. 
The Mandelstam invariants $s_{ij}, s_{ijk} ,  s_{ijkl},\ldots$ which appear in
the BRST variation (\ref{QTijkl}) play a crucial role: They must be the propagators associated with the $T_{ijkl}$ 
to guarantee that each term in $QT_{j_1 \ldots  j_p}$ cancels one of the poles. This is the only way to combine 
different terms $T_{j_1 \ldots  j_p} / (s_{j_1 j_2}, s_{j_1j_2j_3} , \ldots, s_{j_1 \ldots  j_p})$ to an overall 
BRST closed SYM- or superstring amplitude.

The $\l$ ghost number one of the $T_{j_1 \ldots  j_p}$ implies that it just represents a subdiagram with $p$ 
on-shell legs and one off-shell leg. Adding all the color ordered diagrams contributing to a $p+1$ point 
amplitude gives rise to a Berends-Giele current $M_{j_1 \ldots  j_p}$, these objects were firstly considered 
in the context of gluon scattering \cite{Berends:1987me}.

\begin{center}
\begin{tikzpicture} [scale=0.8,line width=0.30mm]
\draw (-2.5,0) node{$M_{12} \ \ = \ \ $};
\draw (0,0) -- (-1,1) node[left]{$k_2$};
\draw (0,0) -- (-1,-1) node[left]{$k_1$};
\draw (0,0) -- (1.7,0);
\draw (1.4,0.3) node{$(k_1+k_2)^2$};
\draw (1.4,-0.3) node{$= s_{12} \neq 0$};
\draw (2.1, 0) node{${\ldots} $};
\draw (3.5,0) node{$\ \ = \ \ \displaystyle {T_{12} \over s_{12}} \ , $};
\draw (8,0) node{$ k_1^2 = k_2^2 = 0$};
\end{tikzpicture}
\end{center}

Let us give explicit lower order examples of $M_{j_1 \ldots  j_p}$ at $p=2,3,4,5$: The $p=2$ 
case $M_{i_1i_2} := T_{i_1i_2}/s_{i_1i_2}$ just represents the cubic vertex of an off-shell 
three--point amplitude. The next examples $p \geq 3$ involve $P_{p+1}=2,5,14,\ldots $ terms according to the color 
ordered $(p+1)$ point amplitudes\footnote{The number $P_{n}$ of pole channels in an $n$ point 
amplitude will be recursively and explicitly given in equation (\ref{poles}) and the line after.}:

\begin{center}
\begin{tikzpicture} [line width=0.30mm]
\draw (-2,0) node{$M_{123} \ \ = $};
\draw (0,0) -- (-1,1) node[above]{$2$};
\draw (0,0) -- (-1,-1) node[below]{$1$};
\draw (0,0) -- (1.8,0);
\draw (0.5,-0.2) node{$s_{12}$};
\draw (1,0) -- (1,1) node[above]{$3$};
\draw (1.5,0.2) node{$s_{123}$};
\draw (2.35, 0) node{${\ldots} \  \ + $};
\draw (8.3,0) node{$\ = \ \ \displaystyle {1 \over s_{123}}  \Bigl( {T_{123} \over s_{12} } + { T_{321} \over s_{23}} \Big)$};
\scope[xshift=-1.5cm]
\draw (5.5,0) -- (4.5,1) node[above]{$3$};
\draw (5.5,0) -- (4.5,-1) node[below]{$2$};
\draw (5.5,0) -- (7.3,0);
\draw (6,-0.2) node{$s_{23}$};
\draw (6.5,0) -- (6.5,-1) node[below]{$1$};
\draw (7,0.2) node{$s_{123}$};
\draw (7.6, 0) node{${\ldots} $};
\endscope
\end{tikzpicture}

\begin{tikzpicture} [scale=1.4,line width=0.30mm]
\draw (-1.5,2) node {$M_{1234} \ \ = $};
\scope[yshift=2cm]
\draw (0,0) -- (-0.5,0.5) node[above]{$2$};
\draw (0,0) -- (-0.5,-0.5) node[below]{$1$};
\draw (0,0) -- (1.3,0);
\draw (0.25,-0.2) node{$s_{12}$};
\draw (0.5,0) -- (0.5,0.5) node[above]{$3$};
\draw (0.75,0.2) node{$s_{123}$};
\draw (1,0) -- (1,0.5) node[above]{$4$};
\draw (1.35,-0.2) node{$s_{1234}$};
\draw (1.5, 0) node{$\ldots  $};
\endscope
\scope[xshift=2.8cm]
\draw (0,2) -- (-0.5,2.5) node[above]{$3$};
\draw (0,2) -- (-0.5,1.5) node[below]{$2$};
\draw (-0.75,2) node{$+$};
\draw (0,2) -- (1.3,2);
\draw (0.25,1.8) node{$s_{23}$};
\draw (0.5,2) -- (0.5,1.5) node[below]{$1$};
\draw (0.75,2.2) node{$s_{123}$};
\draw (1,2) -- (1,2.5) node[above]{$4$};
\draw (1.35,1.8) node{$s_{1234}$};
\draw (1.5,2) node{$\ldots  $};
\endscope
\scope[xshift=1.4cm]
\draw (4.2, 2) -- (3.7, 2.5) node[above]{$4$};
\draw (4.2, 2) -- (3.7, 1.5) node[below]{$3$};
\draw (3.45,2) node{$+$};
\draw (4.2, 2) -- (5.5, 2);
\draw (4.45, 1.8) node{$s_{34}$};
\draw (4.7, 2) -- (4.7, 1.5) node[below]{$2$};
\draw (4.95, 2.2) node{$s_{234}$};
\draw (5.2, 2) -- (5.2, 1.5) node[below]{$1$};
\draw (5.55, 1.8) node{$s_{1234}$};
\draw (5.7, 2) node{$\ldots  $};
\endscope
\scope[xshift=-4.2cm]
\draw (4.2, 0) -- (3.7, 0.5) node[above]{$3$};
\draw (4.2, 0) -- (3.7, -0.5) node[below]{$2$};
\draw (3.45,0) node{$+$};
\draw (4.2, 0) -- (5.5, 0);
\draw (4.45,-0.2) node{$s_{23}$};
\draw (4.7, 0) -- (4.7, 0.5) node[above]{$4$};
\draw (4.95, -0.2) node{$s_{234}$};
\draw (5.2, 0) -- (5.2, -0.5) node[below]{$1$};
\draw (5.45, 0.2) node{$s_{1234}$};
\draw (5.7, 0) node{$\ldots  $};
\endscope
\scope[xshift=1.8cm,yshift=1.5cm]
  \draw (1, -1.5) -- (4.0, -1.5);
  \draw (1, -1.5) -- (0.5, -1.0) node[above]{$2$};
  \draw (1, -1.5) -- (0.5, -2.0) node[below]{$1$};
  \draw (0.25,-1.5) node{$+$};
  \draw (4.0, -1.5) -- (4.5, -1.0) node[above]{$3$};
  \draw (4.0, -1.5) -- (4.5, -2.0) node[below]{$4$};
  \draw (2.5, -1.5) -- (2.5, -1.75);
  \draw (2.5, -1.85) node{$\vdots$};
  \draw (1.75, -1.3) node{$s_{12}$};
  \draw (3.25, -1.3) node{$s_{34}$};
  \draw (2.8, -1.7) node{$s_{1234}$};
\endscope
\draw (2.4,-1.5) node{$\displaystyle = \ \ {1 \over s_{1234}} \Big( {T_{1234}\over s_{12}s_{123} } 
+ {T_{3214}\over s_{23}s_{123} }+ {T_{3421} \over s_{34}s_{234} } + {T_{3241} \over s_{23}s_{234} } 
+ {2T_{12[34]}\over s_{12}s_{34} }\Big)$};
\end{tikzpicture}
\end{center}
According to $P_5 = 5$, there are five diagrams collected in $M_{1234}$
and the last one makes use of the fact that $QT_{12[34]}$
cancels poles in $s_{12},s_{34}$ and $s_{1234}$. As we have mentioned before, the 
diagrammatic interpretation of the BRST building blocks rests on their symmetry 
properties such as $T_{(ij)}= T_{(ij)k}= T_{[ijk]} = 0$ at $p=2,3$. In the $p=4$ 
case at hand, $T_{12[34]} + T_{34[12]} = 0$ is crucial to preserve the reflection 
symmetry $(1,2,3,4) \leftrightarrow (4,3,2,1)$ of the last diagram in the figure above.

As a last explicit example, we shall display $M_{12345}$ here:
\begin{widetext}
\be\label{Ms}
\ba{lcl}
M_{12345} &\equiv&
\ds{ {1 \over s_{12345}}\Big[
        {T_{12345} \over s_{12} s_{123} s_{1234}}
       - {T_{23145} \over s_{23} s_{123} s_{1234}}
       - {T_{23415} \over s_{23} s_{234} s_{1234}}
       + {T_{34215} \over s_{34} s_{234} s_{1234}}
       - {T_{23451} \over s_{23} s_{234} s_{2345}} }\\[3mm]
     && \ds{  + {T_{34251} \over s_{34} s_{234} s_{2345}}
       + {T_{34521} \over s_{34} s_{345} s_{2345}}
       - {T_{45321} \over s_{45} s_{345} s_{2345}}
       + {(T_{34215} - T_{34125}) \over s_{12} s_{34} s_{1234}}
       + {(T_{45231} - T_{45321}) \over s_{23} s_{45} s_{2345}}
       \Big] }\\[3mm]
&&\ds{ + {1\over s_{12345}} \Big[
         {(T_{12345} + T_{21354})\over s_{12} s_{45} s_{123}}
       - {(T_{23145} + T_{32154})\over s_{23} s_{45} s_{123}}
       - {(T_{34512} + T_{43521})\over s_{12} s_{34} s_{345}}
       + {(T_{45312} + T_{54321})\over s_{12} s_{45} s_{345}}
       \Big]\ .}
\ea
\ee
\end{widetext}
The 14 cubic graphs encompassed by $M_{12345}$ as well as higher rank currents can be found 
in an appendix of \cite{WIP}. Apart from this diagrammatic method to construct $M_{i_1 \ldots  i_p}$,
we will give a string-inspired formula in section \ref{SEC4}.

\subsection{Berends-Giele recursions for SYM amplitudes}

Remarkably, the BRST variation of Berends-Giele currents $M_{12{\ldots} p}$ introduces 
bilinears of lower rank $M_{12{\ldots} j<p}$. Up to $p=4$, these are
\be\label{QMijkl}
\ba{lcl}
Q M_{ij} &=& V_{i}V_{j} =: E_{ij} , \ \ \ QM_{ijk} = V_i M_{jk} + M_{ij}V_k =: E_{ijk}
\\
QM_{ijkl} &=& V_i M_{jkl} + M_{ij} M_{kl} + M_{ijk} V_l =: E_{ijkl} \ .
\ea
\ee
More generally, the BRST charge cuts $M_{12{\ldots} p}$ into all color ordered partitions of 
its $p$ on-shell legs among two lower rank Berends-Giele currents
\be\label{Es}
QM_{12{\ldots} p} = \sum_{j=1}^{p-1} M_{12{\ldots} j} M_{j+1{\ldots} p} =: E_{12{\ldots} p}
\ee
where the one-index version is defined to be the unintegrated SYM vertex operator $M_i = V_i$. 
We have explicitly obtained solutions to (\ref{Es}) up to $M_{12{\ldots} 7}$ \cite{WIP}.

Let us denote the number of kinematic poles configurations in $M_{i_1{\ldots} i_p}$ or 
$E_{i_1{\ldots} i_p}$ by $P_{p+1}$, then \req{Es} implies the recursion relation
\be\label{poles}
P_n = \sum_{i=2}^{n-1} P_iP_{n-i+1},\quad P_2 = P_3 \equiv 1,\quad\quad n\ge 4\ .
\ee
Its explicit solution $P_n = 2^{n-2}{(2n-5)!!\over (n-1)!}$ agrees with the formula of  for 
the number of cubic diagrams in the color ordered $n$--point SYM amplitude, see e.g. \cite{BCJ}. 
Hence, our expression ${\cal A}_n = \langle E_{i_1{\ldots} i_{n-1}} V_n \rangle$ passes the 
consistency check to encompass the right number of diagrams.

We have defined the rank $p$ Berends-Giele currents $M_{i_1{\ldots} i_p}$ to contain $p-1$ inverse 
powers of Mandelstam invariants $s_{i_1{\ldots} i_q} = \half (k_{i_1} + {\ldots} + k_{i_q})^2$ and 
in particular an overall propagator $M_{i_1{\ldots} i_p} \sim (s_{i_1{\ldots} i_p})^{-1}$. The latter 
cancels under action (\ref{Es}) of the BRST charge such that the resulting $\l$ ghost number two 
superfield $QM_{i_1{\ldots} i_p} = E_{i_1{\ldots} i_p}$ is well defined even if $s_{i_1{\ldots} i_p}=0$. 

Actually, this is the crucial reason why ${\cal A}_n = \langle E_{i_1{\ldots} i_{n-1}} V_n \rangle$ lies 
in the BRST cohomology: Massless $n$--particle kinematics imply that $s_{i_1{\ldots} i_{n-1}}=0$. The 
resulting rank $n-1$ Berends-Giele current $M_{i_1{\ldots} i_{n-1}}$ diverges due to the overall propagator 
and we cannot write $E_{i_1{\ldots} i_{n-1}}$ as a BRST variation. The $s_{i_1{\ldots} i_{n-1}}=0$ constraint 
saves ${\cal A}_n$ from being BRST exact! Expressing the $n$--point amplitude in terms 
of $E_{i_1{\ldots} i_{n-1}}$ amounts to removing the overall pole before putting the rank $n-1$ Berends-Giele current on-shell.

The representation of the SYM $n$--point amplitude as a bilinear in Berends-Giele currents
\be\label{bilin}
{\cal A}_n = \sum_{j=1}^{n-2} \langle   M_{12{\ldots} j} M_{j+1{\ldots} n-1} V_n \rangle
\ee
makes its factorization into $(j+1)$--point and $(n-j)$--point subamplitudes manifest, see the following figure


\hskip\parindent
\begin{center}
\begin{tikzpicture} [scale=1.0,line width=0.30mm]
\draw (1,0) node{$\displaystyle {\cal A}_n \ \ = \ \ \sum_{j=1}^{n-2}$ };
\draw (4,0) -- (4,1.8) node[above]{$j$};
\draw (4,0) -- (4,-1.8) node[below]{$1$};
\draw (4,0) -- (2.75,-1.25) node[below]{$2$};
\draw[dashed] (4,1.4) arc (90:225:1.4cm);
\draw[fill=white] (4,0) circle (0.7cm);
\draw (4,0) node{$M^j$ };
\draw (4.7,0) -- (6.3,0);
\draw (5.5,0) -- (5.5,-1.4)node[below]{$V_n$};
\draw (7,0) -- (7,1.8) node[above]{$j+1$};
\draw (7,0) -- (8.25,1.25) node[above]{$j+2$};
\draw (7,0) -- (7,-1.8) node[below]{$n-1$};
\draw[dashed] (7,-1.4) arc (-90:45:1.4cm);
\draw[fill=white] (7,0) circle (0.7cm);
\draw (7,0) node{$M^{n-j-1}$ };
\end{tikzpicture}
\end{center}

Equations (\ref{bilin}) and (\ref{Es}) can be viewed as a supersymmetric generalization of Berends-Giele 
recursion relations for gluon amplitudes \cite{Berends:1987me}. As an additional bonus, our 
$M_{12\ldots j}$ do not receive contributions from quartic vertices.

\subsection{BRST equivalent expressions for $\bm{{\cal A}_n}$ and cyclic invariance}

It follows from \req{bilin} that $p=n-2$ is the maximum rank of $M_{i_1{\ldots} i_p}$ appearing in the $n$-point amplitude 
cohomology formula \req{Npts}. However, these terms 
are of the form $\langle M_{i_1{\ldots} i_{n-2} }V_{i_{n-1}} V_{i_n}\rangle$ and can be rewritten
as $\langle E_{i_1{\ldots} i_{n-2} }M_{i_{n-1}i_n}\rangle $ due to $V_{i}V_{j} = E_{ij} = QM_{ij}$ and BRST integration by parts 
\be\label{BRSTintegration}
\langle M_{i_1{\ldots} i_p} E_{i_1{\ldots} i_q}\rangle = \langle E_{i_1{\ldots} i_p} M_{i_1{\ldots} i_q}\rangle\ .
\ee
The decomposition of $E_{i_1{\ldots} i_{n-2}}$ involves at most $M_{i_1{\ldots} i_{n-3}}$, so BRST integration
by parts reduces the maximum rank $p$ of $M_{i_1 \ldots  i_p}$ by one. 
It turns out that the $n$--point cohomology formula \req{Npts} allows enough BRST integrations by parts as
to reduce the maximum rank to $p=[n/2]$. The $[\cdot]$ bracket denotes the Gauss bracket $[x] = \max_{n \in \mathbb Z} n \leq x$ which picks out the nearest integer smaller than or equal to its argument. This yields a more economic expression for 
${\cal A}_n$.

Another benefit of the BRST equivalent ${\cal A}_n$ representation in terms of $M_{i_1 \ldots  i_p}$ 
with $p \leq [n/2]$ lies in the manifest cyclic symmetry. The last leg $V_n$ being singled out
in \req{Npts} obscures the amplitudes' cyclicity. Performing $k$ integrations by parts includes 
$V_n$ into bigger blocks $M_{i_1 \ldots  i_{k+1}}$ such that the $n$'th leg appears on the same 
footing as any other one in the end. We will give examples in the following section \ref{SEC3}.

\section{The $\bm{n}$--point amplitudes up to $\bm{n=10}$}
\label{SEC3}

The three-point amplitude \cite{psf} is trivially reproduced by \req{Npts}
and \req{Es},
\be\label{trpt}
A_3 = \langle E_{12}V_3\rangle = \langle V_1V_2V_3 \rangle\ .
\ee
Similarly, \req{Npts} and \req{Es} reproduce the results of \cite{mafraids,mafrabcj,FTAmps} 
for the four--point amplitude:
\bea
{\cal A}_4 &=& \langle E_{123}V_4\rangle = \langle V_1 M_{23} V_4 \rangle + \langle M_{12}V_3 V_4\rangle\nonumber\\
&=& {1\over s_{23}}\langle V_1 T_{23} V_4\rangle + {1 \over s_{12}}\langle T_{12}V_3V_4 \rangle
\label{fourpt}
\eea
For $n=5$, the formul{\ae} \req{Npts} and \req{Es} lead to:
\begin{widetext}
\bea\label{brsttrick}
{\cal A}_5 &=& \langle E_{1234}V_5\rangle
 =  \langle V_1 M_{234}  V_5 \rangle + \langle M_{12} M_{34} V_5 \rangle
  + \langle M_{123} V_4  V_5\rangle\nnn
&=&  {\langle T_{123}V_{4}V_{5}\rangle\over s_{12} s_{45}}
       - {\langle T_{234}V_{1}V_{5}\rangle\over s_{23} s_{51}}
       + {\langle T_{12}T_{34}V_{5}\rangle\over s_{12} s_{34}}
       - {\langle T_{231}V_{4}V_{5}\rangle\over s_{23} s_{45}}
       + {\langle T_{342}V_{1}V_{5}\rangle\over s_{34} s_{51}}\ .
\eea
As discussed in the previous section, identifying $E_{ij}$ in \req{brsttrick} and using
\req{BRSTintegration} leads to a manifestly cyclic-invariant form proved in \cite{FTAmps}
\be\label{fivec}
{\cal A}_5 = \langle M_{12}V_3M_{45}\rangle + {\rm cyclic}(12345) =
        {\langle T_{12}V_3 T_{45}\rangle\over s_{12} s_{45}} + {\rm cyclic}(12345).
\ee
For $n=6$ the formula \req{Npts} reads
\be\label{sixptmaster}
{\cal A}_6 =   \langle E_{12345}V_6\rangle =
        \langle V_{1} M_{2345}V_{6}\rangle
       + \langle M_{12}M_{345}V_{6}\rangle
       + \langle M_{123}M_{45}V_{6}\rangle
       + \langle M_{1234}V_{5}V_{6}\rangle.
\ee
Integrating the BRST-charge by parts in the first and last terms 
using \req{BRSTintegration} leads to
\be\label{sixptproved}
\ba{lcl}
{\cal A}_6 &=&
        \langle M_{12} M_{34} M_{56} \rangle
       + \langle M_{23} M_{45} M_{61} \rangle
       + \langle M_{123} (M_{45} V_{6} + V_4 M_{56}) \rangle
       + \langle M_{234} ( V_ 5 M_{61} + M_{56} V_{1})\rangle\\[2mm]
&&\ds{+ \langle M_{345} ( V_ 6 M_{12} + M_{61} V_{2})\rangle={\langle T_{12} T_{34} T_{56} \rangle \over 3 s_{12} s_{34} s_{56}}
+ \half \langle \({T_{123}\over s_{12}s_{123}} - {T_{231}\over s_{23} s_{123}}\) \({T_{45} V_6\over s_{45}} + {V_4 T_{56}\over s_{56}}\)\rangle
+ {\rm cyclic}(1{\ldots} 6)\ .}
\ea
\ee
The amplitude \req{sixptproved} was first proposed in \cite{FTAmps} by using BRST cohomology
arguments and proved by the field theory limit of the six-point superstring
amplitude in \cite{MSST}. For $n=7$,
$$
{\cal A}_7 = 
         \langle V_1 M_{23456} V_{7} \rangle
       + \langle M_{12} M_{3456} V_{7} \rangle
       + \langle M_{123} M_{456} V_{7} \rangle
       + \langle M_{1234} M_{56} V_{7} \rangle
       + \langle M_{12345} V_{6} V_{7} \rangle.
$$
Identifying $V_iV_j = E_{ij} = QM_{ij}$ and using \req{BRSTintegration} leads to
$$\ba{lcl}
{\cal A}_7 &=&
         \langle M_{123} M_{45} M_{67} \rangle
       + \langle M_{123} M_{456} V_{7} \rangle
       + \langle M_{234} M_{56} M_{71} \rangle
       + \langle M_{345} M_{67} M_{12} \rangle
       + \langle M_{456} M_{71} M_{23} \rangle\\[2mm]      
       &&+ \langle M_{1234} (V_{5} M_{67} + M_{56} V_{7}) \rangle
       + \langle M_{2345} (V_{6} M_{71} + M_{67} V_{1}) \rangle
       + \langle M_{3456} (V_{7} M_{12} + M_{71} V_{2}) \rangle\ ,
\ea$$
where the generated factors of $E_{12345}$ and $E_{23456}$ 
have been replaced by $M$'s using the definition \req{Es}.
The maximum rank $M_{i_1{\ldots} i_4}$ only appear in combination with
the BRST-exact superfield $E_{ijk} = V_iM_{jk} + M_{ij}V_k = QM_{ijk}$. Using
\req{BRSTintegration} once again leads to a more compact expression with manifest cyclic
symmetry,
\be\label{sevptf}
{\cal A}_7 = \langle M_{123} M_{45} M_{67} \rangle + \langle V_{1} M_{234} M_{567} \rangle
+ {\rm cyclic}(1{\ldots} 7)\ .
\ee
Plugging the solutions \req{Ms} in \req{sevptf} leads to the Ansatz of \cite{FTAmps},
\be\label{sevFT}
{\cal A}_7 =
\langle V_1 \({T_{234}\over s_{23}s_{234}} - {T_{342}\over s_{34}s_{234}}\)\({T_{567}\over s_{56}s_{567}} 
- {T_{675}\over s_{67}s_{567}}\)\rangle
+  \langle \({T_{123}\over s_{12}s_{123}} - {T_{231}\over s_{23}s_{123}}\){T_{45}T_{67}\over  s_{45} s_{67}}\rangle
+ {\rm cyclic}(1{\ldots} 7).
\ee
It is easy to check that \req{sevFT} is expanded in terms of 42 kinematic poles.

The procedure to obtain manifestly cyclic symmetric higher-point amplitudes 
using \req{Npts} and \req{Es} is straightforward and follows the same steps as
above. Increasing the number of legs allows further BRST integrations by parts to
be performed by identifying and integrating $E_{ij}, E_{ijk},{\ldots}$ successively at
each step, leading to
\bea
{\cal A}_8&=& \langle M_{123} M_{456} M_{78}\rangle + \half \langle M_{1234} E_{5678}\rangle 
+ {\rm cyclic}(1{\ldots} 8)\ ,\\
{\cal A}_9 &=& 
{1\over 3}\langle M_{123}M_{456}M_{789}\rangle
+ \langle M_{1234} (M_{567} M_{89} + M_{56}  M_{789} + M_{5678} V_9) \rangle 
+ {\rm cyclic}(1{\ldots} 9)\ ,\\
{\cal A}_{10} &=& \langle M_{1234}(M_{567}M_{89;10} + M_{5678}M_{9;10})\rangle
+\half \langle M_{12345}E_{6789;10}\rangle 
+ {\rm cyclic}(1{\ldots} 10)\ .
\label{nptAmps}
\eea
\end{widetext}

\section{Relation to superstring theory}
\label{SEC4}

Supersymmetric field theory tree--amplitudes can also be obtained from the low--energy limit of superstring theory where the 
dimensionless combinations $\alpha' s_{i_1 \ldots  i_p}$ of Regge slope~$\alpha'$ and Mandelstam bilinears are formally 
sent to zero. Using the pure spinor formalism \cite{psf}, in \cite{WIP,WIPP} the full superstring $n-$point 
amplitude at tree-level is given by
\begin{widetext}
\bea
{\cal A}_n^{{\rm string}}(\alpha') &=&(2\alpha')^{n-3} \prod_{i=2}^{n-2} \int_{z_{i-1}}^{1} \! \! \! \! \! {\rm d} z_i \ 
\prod_{j<k} |z_{jk}|^{-2 \alpha' s_{jk}}  \sum_{p=1}^{n-2} {\langle T_{12\ldots p}  \, T_{n-1,p+1,\ldots ,n-2} \, V_n \rangle 
\over (z_{12} z_{23} \ldots  z_{p-1,p}) (z_{n-1,p+1}  z_{p+1,p+2} \ldots  z_{n-3,n-2})} \notag \\
& +& {\cal P}(2,3,\ldots ,n-2) 
\eea \label{string}
\end{widetext}
where $SL(2,\mathbb R)$ invariance of the tree-level worldsheet admits to fix $(z_1,z_{n-1},z_n)=(0,1,\infty)$ and ${\cal P}(2,3,{\ldots},n-2)$ denotes
a sum over all permutations of $(2,3,\dots,n-2)$.
The full superstring amplitude is determined by BRST building blocks $T_{12\ldots p}$ and $n-3$ worldsheet integrals over $z_{jk}=z_j-z_k$. The $\alpha' \rightarrow 0$ 
limit of \req{string} reproduces ${\cal A}_n = \sum\limits_{p=1}^{n-2} \langle M_{i_1{\ldots} i_p} M_{i_{p+1}{\ldots} i_{n-1}} V_n \rangle$ 
term by term in the individual $p$ sums. Therefore considering $p=n-2\equiv q$ yields an explicit 
formula for $M_{i_1{\ldots} i_p}$
\begin{widetext}
\be\label{Mstring}
M_{12\ldots q} = \lim_{\alpha' \rightarrow 0} (2\alpha')^{q-1} \prod_{i=2}^{q} \int_{z_{i-1}}^{1} \! \! \! \! \! {\rm d} z_i \ 
\prod_{j<k}^{q+1} |z_{jk}|^{-2 \alpha' s_{jk}} \( {T_{12\ldots q} \over z_{12} z_{23} \ldots  z_{q-1,q} } 
+ {\cal P}(2,3,\ldots ,q) \)
\ee
\end{widetext}
in the fixing $z_1=0$ and $z_{q+1}=1$. It has been checked up to $q=7$ that the string inspired computation
\req{Mstring} of $M_{12\ldots q}$ agrees with its construction from the color ordered diagrams in ${\cal A}_{q+1}$.

\vskip0.5cm
{\bf Acknowledgments:}
CRM acknowledges support by the Deutsch-Israelische Projektkooperation (DIP H52)
and thanks the Werner--Heisenberg--Institut in M\"unchen for hospitality and
partial financial support.
OS would like to thank Stefan Theisen and the  Albert--Einstein--Institut in
Potsdam for warm 
hospitality and generous support during the time of writing.
St.St. would like to thank the Albert--Einstein--Institut in
Potsdam and in particular
Hermann Nicolai and Stefan Theisen for invitation and partial support
during preparation of this work.

\nocite{*}
\bibliography{msst0706}
\bibliographystyle{h-physrev5}

\end{document}